\documentclass[conference]{IEEEtran}
\IEEEoverridecommandlockouts
\usepackage[pdftex]{graphicx}
\graphicspath{{BlockDiagram}{SimulationResults}}
\usepackage{cite,hyperref}
\usepackage{float}
\usepackage{booktabs}
\usepackage{amsmath,amssymb,amsfonts}
\usepackage{algorithmic}
\usepackage{graphicx}
\usepackage{subfigure}
\usepackage{textcomp}
\usepackage{xcolor}
\usepackage{pbox}
\usepackage{stfloats}
\usepackage{cuted} \stripsep +0pt 

\def\BibTeX{{\rm B\kern-.05em{\sc i\kern-.025em b}\kern-.08em
    T\kern-.1667em\lower.7ex\hbox{E}\kern-.125emX}}

\begin{document}

	\title{Auto-Optimized Maximum Torque Per Ampere Control of IPMSM Using Dual Control for Exploration and Exploitation 
		\\
	\thanks{
	        This work was supported in part by the UK Engineering and Physical Science Research Council (EPSRC) New Investigator Award under Grant EP/W027283/1.
	        Dr. Jun Yang is the corresponding author. Emails: \{y.zuo, y.yu2, j.yang3, w.chen\}@lboro.ac.uk
	    }
	}
	
	\author{
		\IEEEauthorblockN{Yuefei Zuo, Yalei Yu, Jun Yang, Wen-Hua Chen}\\
		\IEEEauthorblockA{\textit{Department of Aeronautical and Automotive Engineering, Loughborough University, United Kingdom}}
	 }

\maketitle

\begin{abstract}
	In this paper, a maximum torque per ampere (MTPA) control strategy for the interior permanent magnet synchronous motor (IPMSM) using dual control for exploration and exploitation (DCEE). In the proposed method, the permanent magnet flux and the difference between the $d$- and $q$-axis inductance are identified by multiple estimators using the recursive least square method. The initial values of the estimated parameters in different estimators are different. By using multiple estimators, exploration of the operational environment to reduce knowledge uncertainty can be realized. Compared to those MTPA control strategies based on the extremum-seeking method, the proposed method has better dynamic performance when speed or load varies. The effectiveness of the proposed method is verified by simulations.
\end{abstract}


\section{Introduction}

	Due to its high power density, high speed range and nice flux-weakening performance, the interior permanent-magnet synchronous motor (IPMSM) has been widely implemented in electric vehicle applications \cite{cai2022critical}.  To achieve the minimum copper loss, the maximum torque per ampere (MTPA)  control strategy is employed to control the current vector under different load conditions. However, MTPA control of IPMSM is challenging due to the variable motor parameters caused by magnetic saturation, temperature, and cross-coupling effects \cite{rang2004mtpa}. 
	
	Existing MTPA methods can be categorised into two types: offline methods and online methods. In off-line methods, motor parameters under different operating conditions are tested, then lookup table (LUT) \cite{jung2013current} or curve fitting techniques \cite{morimoto1994effects} are used to vary motor parameters with operating conditions. However, these offline methods are time-consuming and some factors, such as manufacturing imperfection and temperature variations, have to be neglected for simplification. To this end, online methods have been developed and have gained more attraction. Among many online methods, signal injection methods based on extremum seeking \cite{bolognani2010online} are widely used due to its effectiveness and easy implementation.  
	
	The principle of extremum seeking can be briefly described as follows. Injecting high-frequency current or current vector angle to generate high-frequency torque,  the gradient of torque over current vector angle can be extracted and forced to zero by using an integrator or proportional and integral (PI) regulator, the output of PI regulator is set as the optimal current vector angle. It should be noted that the real signal injection causes torque ripple, which degrades the torque control performance. To solve this problem, the virtual signal injection (VSI) method is proposed in \cite{sun2015maximum}. The VSI method is also based on extremum seeking principle, but it does not generate torque ripple since the torque used for gradient extraction is calculated by a model which uses voltage and speed instead of the flux linkage. The accuracy of the VSI method depends on the accuracy of the model for torque calculation \cite{sun2016accuracy}, it can be improved by considering derivative terms during the dynamic process \cite{sun2021extended}. In extremum seeking methods, a band-pass filter (BPF) and a low-pass filter are generally required. These two filters limit the improvement of the dynamic performance of the system and cause the system to oscillate in the dynamic process. This problem can be solved by using square-wave injection \cite{zhao2016virtual}, which is widely used in sensorless control. Though extermum seeking methods are able to achieve accurate MTPA control in the steady state, they may not realise accurate MTPA control in the transient state because it takes some time to find the new optimal point.
	
	In this paper, a MTPA control strategy using the dual control for exploration and exploitation (DCEE) \cite{chen2021dual,li2022concurrent} is proposed. To calculate the MTPA operating point, the proposed method identifies the permanent magnet flux and the difference between the $d$- and $q$-axis inductance using the recursive least square method. More importantly, multiple estimators with different initial values of estimated parameters are employed and DCEE algorithm is applied to generate better control output for estimating parameters. Simulation results show that the proposed MTPA control strategy has better dynamic performance than the extremum-seeking-based MTPA control strategies.

\section{Problem Formulation}

    In this section, we will introduce the basic idea of using MTPA control to realize copper loss minimization of IPMSM.

	\subsection{Mathematical Model of IPMSM}

		The motion equation of IPMSM can be expressed as

        \begin{equation} \label{eq:Motion equation}
			T_e=J\frac{d\omega _m}{dt}+B\omega _m+T_L
		\end{equation}

        \noindent where $T_e$ and $T_L$ are the electromagnetic torque generated by the IPMSM and the load torque, $J$ and $B$ denote the inertia and the viscous friction torque coefficient, and $\omega _m$ represents the mechanical angular speed of the rotor.

        To realize speed control of IPMSM, the electromagnetic torque $T_e$ consisting of the permanent magnet torque and the reluctance torque \cite{jahns1986interior} should be regulated.

        \begin{equation} \label{eq:Torque equation}
			T_e=\underset{\mathrm{Permanent Magnet Torque}}{\underbrace{1.5p_n\psi _fi_q}}+\underset{\mathrm{Reluctance Torque}}{\underbrace{1.5p_n\left( L_d-L_q \right) i_di_q}}
		\end{equation}

        \noindent where $p_n$, $\psi_f$, $L_d$, and $L_q$ are motor parameters that decided when the motor is designed, represent the pole pair numbers, the permanent magnet flux linkage, and the $d$- and $q$-axis inductance, respectively; $i_d$ and $i_q$ are the $d$- and $q$-axis currents.

        The two currents $i_d$ and $i_q$ can be regulated by changing the $d$- and $q$-axis voltages $u_d$ and $u_q$, which are given by

		\begin{equation} \label{eq:Voltage equations}
            \begin{cases}
                u_d=R_si_d+L_d\frac{di_d}{dt}-\omega _rL_qi_q \\
                u_q=R_si_q+L_q\frac{di_q}{dt}+\omega _r\left( L_di_d+\psi _f \right) \\
            \end{cases}			
		\end{equation}

        \noindent where $\omega _r=p_n\omega _m$ is the electrical angular speed of the rotor.

        The copper loss $P_\mathrm{cu}$ is the power loss caused by the currents going through the resistor, is given by

        \begin{equation} \label{eq:Copper Loss}
			P_\mathrm{cu}=3R_s i_{s}^{2}=3R_s\left( i_{d}^{2}+i_{q}^{2} \right) 
		\end{equation}

        \noindent where $i_s=\sqrt{i_{d}^{2}+i_{q}^{2}}$ is the amplitude of the phase current.      

    \subsection{Principle of MTPA Control}    

\begin{figure} [!t]
        	\centering
        	\setlength{\abovecaptionskip}{0cm}	
        	\includegraphics[width=0.8\columnwidth]{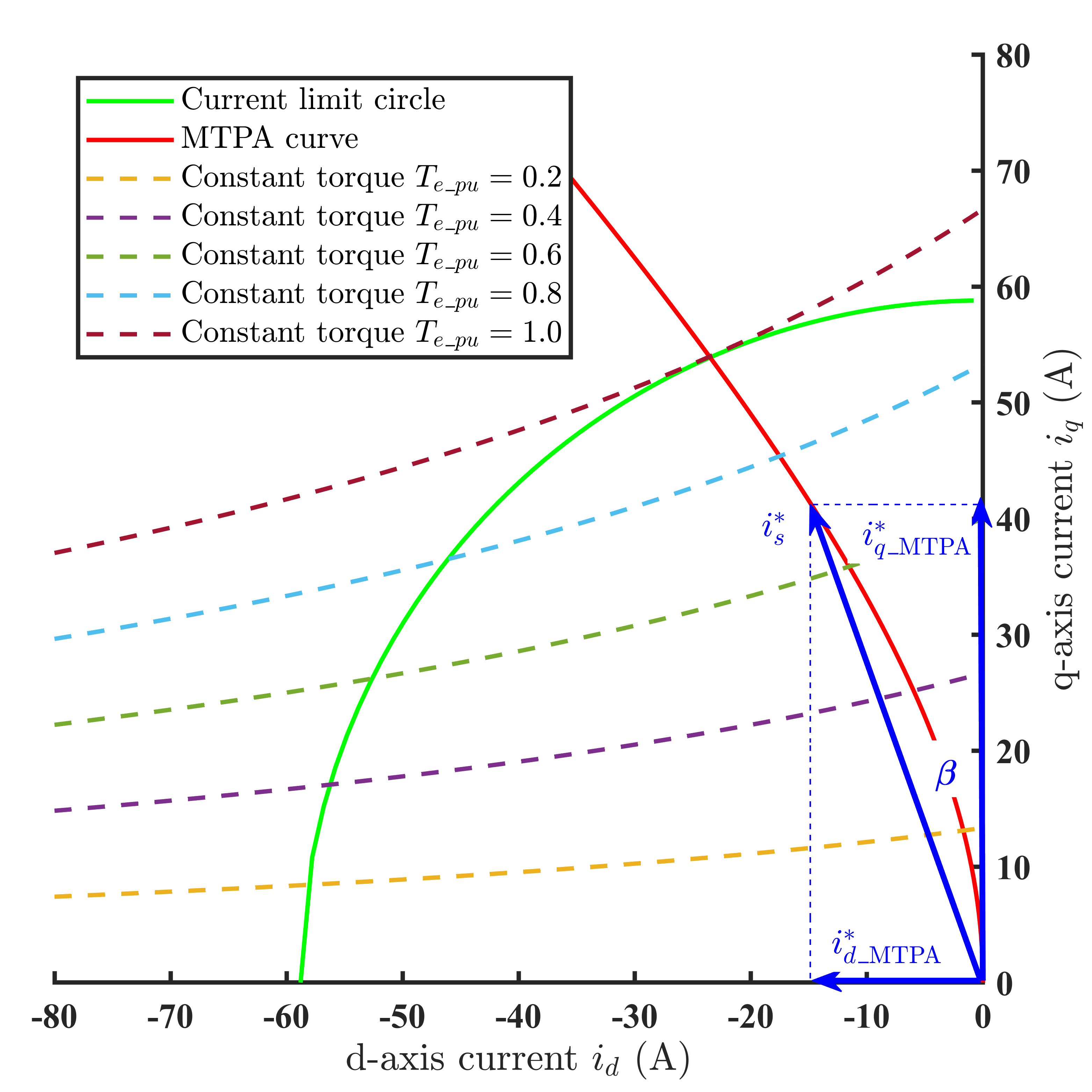}
        	\caption{Current Trajectory}
        	\label{fig:MTPA_Curve}       
        	\vspace{-3ex} 	
        \end{figure}

        Generally, $\omega _m$ is constant in the steady state and $T_e=B\omega _m+T_L$ is also constant. From \eqref{eq:Torque equation}, it is known that both $i_d$ and $i_q$ contribute to the torque and there are infinite combinations of $i_d$ and $i_q$ to generate the same torque. So, the operating point locates at the constant torque curve. To minimize the copper loss, the motor should work at the point where $i_s$ is the minimum, i.e., MTPA point. By connecting all the MTPA points under different torque $T_e$, the MTPA curve can be obtained, as shown in Fig. \ref{fig:MTPA_Curve}. In Fig. \ref{fig:MTPA_Curve}, the torque per unit value $T_{e\_pu}$ is the ratio of $T_e$ over the rated torque $T_N$, i.e., $T_{e\_pu}=T_e/T_N$.

\begin{figure} [!t]
        	\centering
        	\setlength{\abovecaptionskip}{0cm}	
        	\includegraphics[width=1.0\columnwidth]{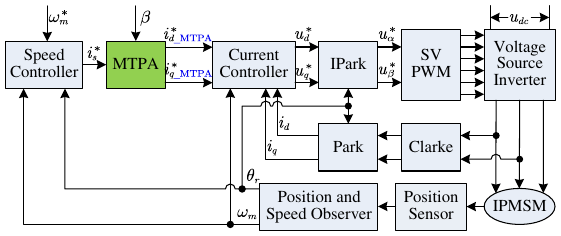}
        	\caption{Block diagram of the IPMSM servo control system}
        	\label{fig:Servo control system}       
        	\vspace{-3ex} 	
        \end{figure}

		In real applications, it is difficult to accurately express $T_e$ as a polynomial of $i_d$ and $i_q$. To this end, the current reference $i_s^*$ is used instead of the torque reference $T_e^*$ as the output of the speed controller \cite{morimoto1994effects}, as shown in Fig. \ref{fig:Servo control system}. For the MTPA curve, with the current vector angle $\beta$, the two currents $i_{d\_\mathrm{MTPA}}^*$ and $i_{q\_\mathrm{MTPA}}^*$ can be expressed as

		\begin{equation}	\label{eq:id and iq for MTPA }
			\begin{cases}
            	i_{base}=\dfrac{\psi _f}{L_q-L_d}\\
            	\beta =\sin ^{-1}\left\{ \dfrac{\sqrt{i_{base}^{2}+8i_{s}^{*2}}-i_{base}}{4i_s} \right\}\\
            	i_{d\_\mathrm{MTPA}}^*=-i_s^*\sin \beta\\
            	i_{q\_\mathrm{MTPA}}^*=i_s^*\cos \beta\\
            \end{cases}		
		\end{equation}

		It can be seen from \eqref{eq:id and iq for MTPA } that the optimal $d$-axis current to realize MTPA control is decided by the two parameters $\psi_f$ and $L_q-L_d$. However, it is challenging to realize MTPA control in real applications because $\psi_f$, $L_d$, and $L_q$ may vary with currents and temperature due to the effect of magnetic circuit saturation \cite{sun2015maximum}.



\section{Proposed MTPA Control Using DCEE}

        In this section, the proposed method of using the dual control for exploration and exploitation (DCEE) concept to realize MTPA control of IPMSM is explained.

    \subsection{Dual Control Reformulation}

        As the pole pair number $p_n$ is a constant, the torque equation expressed by \eqref{eq:Torque equation} can be modified as

		\begin{equation}	\label{eq:Modified torque equation}		
            \begin{aligned}
                T_{e1}&=\frac{2T_e}{3p_n}=i_q\psi _f-i_di_q\left( L_q-L_d \right) 
                \\
                &=\left[ \begin{matrix}
                	i_q&		-i_di_q\\
                \end{matrix} \right] \left[ \begin{array}{c}
                	\psi _f\\
                	L_q-L_d\\
                \end{array} \right] 
                \\
                &=\boldsymbol{\phi }^\mathrm{T}\left( i_d,i_q \right) \boldsymbol{\theta }
            \end{aligned}			
		\end{equation}
		
		\noindent where 
		$
		\boldsymbol{\phi }\left( i_d,i_q \right) =\left[ \begin{matrix}
        	i_q&		-i_di_q\\
        \end{matrix} \right] ^\mathrm{T}
		$ is the basis function, and
		$
		\boldsymbol{\theta }=\left[ \begin{matrix}
        	\psi _f&		L_q-L_d\\
        \end{matrix} \right] ^\mathrm{T}
		$ is the unknown motor parameter.

        The system dynamics under concern are described by

        \begin{equation} \label{eq:State equation of current}
            \left[ \begin{array}{c}
            	\dot{i}_d\\
            	\dot{i}_q\\
            \end{array} \right] =\left[ \begin{matrix}
            	-\frac{R_s}{L_d}&		\omega _r\frac{L_q}{L_d}\\
            	-\omega _r\frac{L_d}{L_q}&		-\frac{R_s}{L_q}\\
            \end{matrix} \right] \left[ \begin{array}{c}
            	i_d\\
            	i_q\\
            \end{array} \right] +\left[ \begin{matrix}
            	\frac{1}{L_d}&		0\\
            	0&		\frac{1}{L_q}\\
            \end{matrix} \right] \left[ \begin{array}{c}
            	u_d\\
            	u_{q1}\\
            \end{array} \right] 
        \end{equation}

        \noindent where $u_{q1}=u_q-\omega _r\psi _f$ is the equivalent control in $q$-axis, both states $i_d$ and $i_q$ can be measured.

        Rewrite the system dynamics in discrete-time domain as

        \begin{equation} \label{eq:Discrete-time state equation of current}
            \boldsymbol{x}\left( k+1 \right) =\left( \boldsymbol{I}+\boldsymbol{A} \right)\boldsymbol{x}\left( k \right) +\boldsymbol{Bu}\left( k \right) 
        \end{equation}

        \noindent where $\boldsymbol{x}=[i_d, i_q]^\mathrm{T}$, $\boldsymbol{u}=[u_d, u_{q1}]^\mathrm{T}$. Matrices $\boldsymbol{A}$ and $\boldsymbol{B}$ can be expressed as
        $$
            \boldsymbol{A}=T_s\left[ \begin{matrix}
                                    	-\dfrac{R_s}{L_d}&		\omega _r\dfrac{L_q}{L_d}\\
                                    	-\omega _r\dfrac{L_d}{L_q}&		-\dfrac{R_s}{L_q}\\
                                    \end{matrix} \right] , \boldsymbol{B}=T_s\left[ \begin{matrix}
                                    	\dfrac{1}{L_d}&		0\\
                                    	0&		\dfrac{1}{L_q}\\
                                    \end{matrix} \right] 
                                    $$ 
        where $T_s$ is the sampling time.

        Since the optimal operation condition $\boldsymbol{r}^*=[i_{d\_\mathrm{MTPA}}^*, i_{q\_\mathrm{MTPA}}^*]^\mathrm{T}$ is unknown, where $ i_{d\_\mathrm{MTPA}}^* $ and $ i_{q\_\mathrm{MTPA}}^* $ are given by (\ref{eq:id and iq for MTPA }), the best one can do is to drive the system to the best estimation of the optimal operation condition, denoted by $ \bar{\boldsymbol{r}}(k+1|k) $, with all the information up to now. This can be formulated as

        \begin{equation} \label{eq:Dual control formulation}
            \underset{u\left( k \right)}{\min}\mathbb{E} \left\{ \left[ \boldsymbol{x}\left( k+1|k \right) -\boldsymbol{r}^* \right] ^\mathrm{T}\left[ \boldsymbol{x}\left( k+1|k \right) -\boldsymbol{r}^* \right] | \boldsymbol{\mathcal{I}}(k+1|k) \right\} 
        \end{equation}
        where $ \boldsymbol{\mathcal{I}}(k+1|k)=\{I_0, I_1, \cdots, I_k, I_{k+1} \}$ denotes measurement information collected up to the current time step $k$. 
        Note that $I_k=\{x(k), u(k-1), T_{e1}(k) \}$ denotes information state, and $I_0=\{x(0), T_{e1}(0) \}$. 

        From equation (\ref{eq:id and iq for MTPA }), we know $\boldsymbol{r}^*$ can be regarded as a function of optimal parameter $ \boldsymbol{\theta} $, given by $\boldsymbol{r}^*=f(\boldsymbol{\theta})$, where $\boldsymbol{\theta}= \left [ \begin{matrix} \psi_f & L_q-L_d \end{matrix} \right] $. Define the best estimation of the optimal operation $\bar{\boldsymbol{r}}(k+1|k)$ as 
        \begin{equation}\label{eq:r_bar}
            \bar{\boldsymbol{r}}(k+1|k) = \mathbb{E}\left[ \boldsymbol{r}(k+1|k) | \boldsymbol{\mathcal{I}}(k+1|k) \right]
        \end{equation}
        Thus the estimation error conditional on $\boldsymbol{\mathcal{I}}(k+1|k)$ can be given as 
        \begin{equation}\label{eq:r_tilde}
            \widetilde{\boldsymbol{r}} = \boldsymbol{r}^* - \bar{\boldsymbol{r}}(k+1|k) 
        \end{equation}
        
        Substituting (\ref{eq:r_tilde}) and (\ref{eq:r_bar}) into (\ref{eq:Dual control formulation}) has 
        \begin{align}\label{eq:Dual control formulation derivation}
            & \mathbb{E}\left[ \| \boldsymbol{x}(k+1|k) - \bar{\boldsymbol{r}}(k+1|k) - \widetilde{\boldsymbol{r}} \| | \boldsymbol{\mathcal{I}}(k+1|k) \right] \nonumber \\ 
            = & \mathbb{E}\left[ \|\boldsymbol{x}(k+1|k)- \bar{\boldsymbol{r}}(k+1|k)\|^2 |\boldsymbol{\mathcal{I}}(k+1|k  \right] \nonumber \\
            & - 2 \mathbb{E} \left[  (\boldsymbol{x}(k+1|k)-\bar{\boldsymbol{r}}(k+1|k))^\mathrm{T} \widetilde{\boldsymbol{r}} | \boldsymbol{\mathcal{I}}(k+1|k) \right] \nonumber \\
            & + \mathbb{E}[\| \widetilde{\boldsymbol{r}}  \|^2 | \boldsymbol{\mathcal{I}}(k+1|k)   ]
        \end{align}

        Given that the definition of $  \widetilde{\boldsymbol{r}} $ in (\ref{eq:r_tilde}), and then we have $ \mathbb{E}\left[  \widetilde{\boldsymbol{r}}| \boldsymbol{\mathcal{I}}(k+1|k) \right] = 0 $. Moreover, considering $ \boldsymbol{x}(k+1|k) $ and $ \bar{\boldsymbol{r}}(k+1|k) $ are deterministic, we have $ \mathbb{E} \left[  (\boldsymbol{x}(k+1|k)-\bar{\boldsymbol{r}}(k+1|k))^\mathrm{T} \widetilde{\boldsymbol{r}} | \boldsymbol{\mathcal{I}}(k+1|k) \right] = 0  $. Therefore, the cost function of DCEE can be rewritten as
        \begin{align}\label{eq:cost function of dual control formulation}
            \boldsymbol{D}(\boldsymbol{u}(k)):= & \mathbb{E}\left[ \|\boldsymbol{x}(k+1|k)- \bar{\boldsymbol{r}}(k+1|k)\|^2 |\boldsymbol{\mathcal{I}}(k+1|k  \right] \nonumber \\
            & + \mathbb{E}[\| \widetilde{\boldsymbol{r}}  \|^2 | \boldsymbol{\mathcal{I}}(k+1|k)   ]
        \end{align}
       where the first term in (\ref{eq:cost function of dual control formulation}) denotes the exploitation tracking the optimal operation and the second term denotes the exploration estimating the optimal operation.

	\subsection{Identification of Electrical Parameters Using Forgetting Factor RLS}
		
		Here, we use the forgetting factor recursive least square (RLS) \cite{guo1993performance} method to identify the two parameters, and the algorithm is expressed by

        \begin{equation} \label{eq:Algorithm of RLS}
			\begin{cases}
            	\boldsymbol{P}{\left( k \right)}=\dfrac{\boldsymbol{P}_{\left( k-1 \right)}}{\lambda +\boldsymbol{\phi }{\left( k \right)}^{T}\boldsymbol{P}_{\left( k-1 \right)}\boldsymbol{\phi }{\left( k \right)}}\\
            	\boldsymbol{K}{\left( k \right)}=\boldsymbol{P}{\left( k \right)}\boldsymbol{\phi }{\left( k \right)}=\dfrac{\boldsymbol{P}_{\left( k-1 \right)}\boldsymbol{\phi }{\left( k \right)}}{\lambda +\boldsymbol{\phi }{\left( k \right)}^{T}\boldsymbol{P}_{\left( k-1 \right)}\boldsymbol{\phi }{\left( k \right)}}\\
            	\hat{\boldsymbol{\theta}}_{j\left( k \right)}=\hat{\boldsymbol{\theta}}_{j}\left( k-1 \right)+\boldsymbol{K}{\left( k \right)}\left( T_{e1}\left( k \right)-\boldsymbol{\phi }{\left( k \right)}^{T}\hat{\boldsymbol{\theta}}_{j}\left( k-1 \right) \right)\\
            \end{cases}	
		\end{equation}

        \noindent where $\lambda$ is the forgetting factor, $T_{e1}(k)$ denotes the measured value of $T_{e1}$ at the time $k$, $j=1,2,\cdots , N$ represents the $j^\mathrm{th}$ estimator in the overall $N$ estimators.

        With the identified parameters $\hat{\boldsymbol{\theta}}_{j}(k)$, the estimated optimal reference $\hat{\boldsymbol{r}}_{j}(k)=[\hat{i}_{jd\_\mathrm{MTPA}}^{*}\left( k \right),\hat{i}_{jq\_\mathrm{MTPA}}^{*}\left( k \right)]^\mathrm{T}$ can be calculated by \eqref{eq:Optimal reference at k}

        \begin{equation} \label{eq:Optimal reference at k}
            \begin{cases}
            	\hat{i}_{base}\left( k \right)=\dfrac{\hat{\theta}_{1j}\left( k \right)}{\hat{\theta}_{2j}\left( k \right)}\\
            	\hat{\beta}{\left( k \right)}=\sin ^{-1}\left\{ \dfrac{\sqrt{\hat{i}_{base}^{2}\left( k \right)+8i_{s}^{*2}\left( k \right)}-\hat{i}_{base}\left( k \right)}{4i_{s}\left( k \right)} \right\}\\
            	\hat{i}_{d\_\mathrm{MTPA}}^{*}\left( k \right)=-i_{s}^{*}\left( k \right)\sin \beta {\left( k \right)}\\
            	\hat{i}_{q\_\mathrm{MTPA}}^{*}\left( k \right)=i_{s}^{*}\left( k \right)\cos \beta {\left( k \right)}\\
            \end{cases}
        \end{equation}
        
        The mean value of $\hat{\boldsymbol{r}}_j\left( k \right)$ can be calculated by 
        
        \begin{equation}
            \bar{\boldsymbol{r}}_j\left( k \right) =\frac{1}{N}\sum_{j=1}^N{\hat{\boldsymbol{r}}_j\left( k \right)}
        \end{equation}
        
        \noindent then the objective can be obtained as

        \begin{equation} \label{eq:Objective at k}
            \boldsymbol{D}{\left( k \right)}=\left( \boldsymbol{x}{\left( k \right)}-\bar{\boldsymbol{r}}_{j}\left( k \right) \right) ^2+\frac{1}{N}\sum_{j=1}^N{\left( \bar{\boldsymbol{r}}_{j}\left( k \right)-\hat{\boldsymbol{r}}_{j}\left( k \right) \right) ^2}
        \end{equation}


    \subsection{Prediction of the Objective with One-Step Forward}
  
        Suppose the states at time $k+1$ can be expressed by $\boldsymbol{x}{\left( k+1|k \right)}=\boldsymbol{x}{\left( k \right)}+\Delta \boldsymbol{x}$, where $\Delta \boldsymbol{x}$ is the incremental currents, then the torque at the time $k+1$ estimated by the $j^\mathrm{th}$ estimator can be calculated by

        \begin{equation}
            \hat{T}_{e1\_j}\left( k+1|k \right)=\boldsymbol{\phi }^{\mathrm{T}}\left( x{\left( k+1|k \right)} \right) \hat{\boldsymbol{\theta}}_{j}\left( k \right)
        \end{equation}
        
        Since the measured torque at the time $k+1$ is not available, it is estimated by using the mean value of the estimated torque        
        
		\begin{equation}
			T_{e1}\left( k+1|k \right)=\frac{1}{N}\sum_{j=1}^N{\hat{T}_{e1\_j}\left( k+1|k \right)}				
		\end{equation}
		
		\noindent then the estimated parameters $\hat{\boldsymbol{\theta}}_{j}(k+1|k)$, the estimated optimal reference $\hat{\boldsymbol{r}}_{j}(k+1|k)$ and its mean value $\bar{\boldsymbol{r}}_{j}(k+1|k)$, and the objective $D(k+1|k)$ can be calculated by \eqref{eq:Algorithm of RLS}, \eqref{eq:Optimal reference at k}, and \eqref{eq:Objective at k}.

        Finally, we can calculate the gradient of the objective over the incremental states as

        \begin{equation}
            \frac{\partial D}{\partial \boldsymbol{x}}=\frac{D{\left( k+1|k \right)}-D{\left( k \right)}}{\Delta \boldsymbol{x}}
        \end{equation}

    \subsection{Deduction of the Control Law}
    
		To minimize the objective, the adaptive law of the states can be designed as

        \begin{equation}
            \boldsymbol{x}\left( k+1 \right) =\boldsymbol{x}\left( k \right) -k_x\frac{\partial D}{\partial \boldsymbol{x}}
        \end{equation}

        \noindent where $k_x>0$ is the adaptive gain. 
        
        According to the state equation expressed by \eqref{eq:Discrete-time state equation of current}, the control output can be deduced as

        \begin{equation}
            \boldsymbol{u}\left( k \right) =-\boldsymbol{B}^{-1}\left[ \boldsymbol{Ax}\left( k \right) +k_x\frac{\partial D}{\partial \boldsymbol{x}} \right]  
        \end{equation}
  
\begin{figure} [!t]
			\centering
			\setlength{\abovecaptionskip}{0cm}
			\includegraphics[width=1.0\columnwidth]{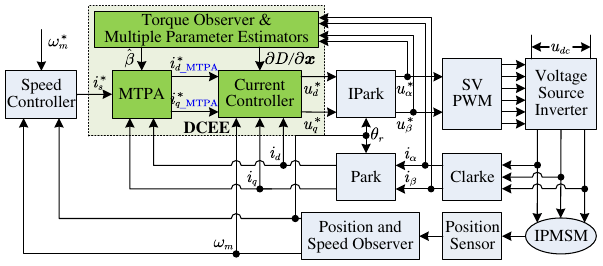}
			\caption{Block diagram of the MTPA control system using DCEE}
			\label{fig:Block diagram of the MTPA-DCEE system}
			\vspace{-3ex}
		\end{figure}

\begin{figure} [!t]
			\centering
			\setlength{\abovecaptionskip}{0cm}
			\includegraphics[width=0.8\columnwidth]{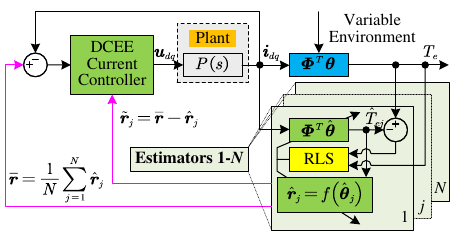}
			\caption{Block diagram of the DCEE controller}
			\label{fig:Block diagram of the DCEE}
			\vspace{-3ex}
		\end{figure}

        The block diagram of the MTPA control system using DCEE is shown in Fig. \ref{fig:Block diagram of the MTPA-DCEE system},
        and the block diagram of the DCEE controller is shown in Fig. \ref{fig:Block diagram of the DCEE}. 

\section{Simulation Results and Analysis}

		In this part, MTPA control schemes based on extremum seeking mentioned in \cite{sun2015maximum} and the proposed DCEE method are compared. The parameters of the IPMSM tested are shown in TABLE \ref{tab_Motor Parameters}. The simulation is carried out in the discrete-time domain with a frequency of 1 MHz, and the sampling time is set as 0.1 ms. An ideal inverter is used in the simulation.
	
		The simulation time is set as 1.0s, rated speed (3000 r/min) is applied from 0s-0.8s and half speed (1500 r/min) is applied from 0.8s-1.0s; no load is applied from 0s-0.2s, full load (36Nm) is applied from 0.2s-0.6s, and half load (18Nm) is applied from 0.6s-1.0s; $i_d=0$ control strategy is adopted from 0s-0.4s, the proposed MTPA control strategy is enabled at 0.4s and adopted from 0.4s-1.0s. To avoid the interrupt change of the load, the step load is smoothed by a S function \cite{zuo2021two}. Through the above setting, five operating conditions are tested, as shown in Table. \ref{tab_Five operating conditions}.

        In simulations, the ideal torque can be obtained directly from the output of the motor model. However, in real applications, it is a common practice to use flux and torque observer \cite{zuo2021digital} to observe the torque. In this paper, both situations are tested.        
	
		\begin{table}[!t]
			\renewcommand{\arraystretch}{1.2}
		    \caption{Parameters of the tested motors}
		    \centering
		    \label{tab_Motor Parameters}
		    \resizebox{\columnwidth}{!}{
			    \begin{tabular}{l l || l l}
				    \hline\hline \\[-3mm]
				    \multicolumn{1}{c}{Parameter} & \multicolumn{1}{c}{Quantity} & 
				    \multicolumn{1}{c}{Parameter} & \multicolumn{1}{c}{Quantity}  \\[0ex] \hline
				    Rated power         $ P_N $     & 10 (kW)      & Polepairs       		$ p_n $     & 3 \\
				    DC voltage       	$ U_{dc} $  & 310 (V)      & d-axis inductance      $ L_d $     & 0.8 (mH) \\
				    Rated speed         $ n_N $     & 3000 (r/min) & q-axis inductance      $ L_q $     & 2.0 (mH) \\
				    Rated torque        $ T_N $     & 36 (Nm)      & Permanent magnet flux  $ \psi_f $  & 0.12 (Wb) \\
				    Current limit       $ I_{smax}$ & 120 (A)      & Stator resistance   	$ R_s $     & 0.05 (Ohm) \\  
				    \hline\hline				    
			\end{tabular}
			}
			\vspace{-3ex}
		\end{table}
		
		\begin{table}[!t]
			
			\renewcommand{\arraystretch}{1.2}
			\caption{Five operating conditions}
			\centering
			\label{tab_Five operating conditions}
			\resizebox{0.6\columnwidth}{!}{
				\begin{tabular}{l l l}
					\hline\hline \\[-3mm]
					\multicolumn{1}{c}{ } 	& \multicolumn{1}{c}{Time range} 	& \multicolumn{1}{c}{Operating conditions}			\\[-0mm] \hline
					1 						& 0s-0.2s 							& $i_d=0$, no load, rated speed 	\\
					2 						& 0.2s-0.4s 						& $i_d=0$, full load, rated speed \\    
					3						& 0.4s-0.6s							& MTPA, full load, rated speed	\\
					4						& 0.6s-0.8s							& MTPA, half load, rated speed	\\
					5						& 0.8s-1.0s							& MTPA, half load, half speed	\\
					\hline\hline
				\end{tabular}
			}
			\vspace{-3ex}
		\end{table}

	\subsection{MTPA Using Extremum Seeking}
	
\begin{figure} [htbp]			
			\centering
                \vspace{-3ex}
			\setlength{\abovecaptionskip}{-1ex}	
			\subfigure[]{\includegraphics[width=0.86\columnwidth]{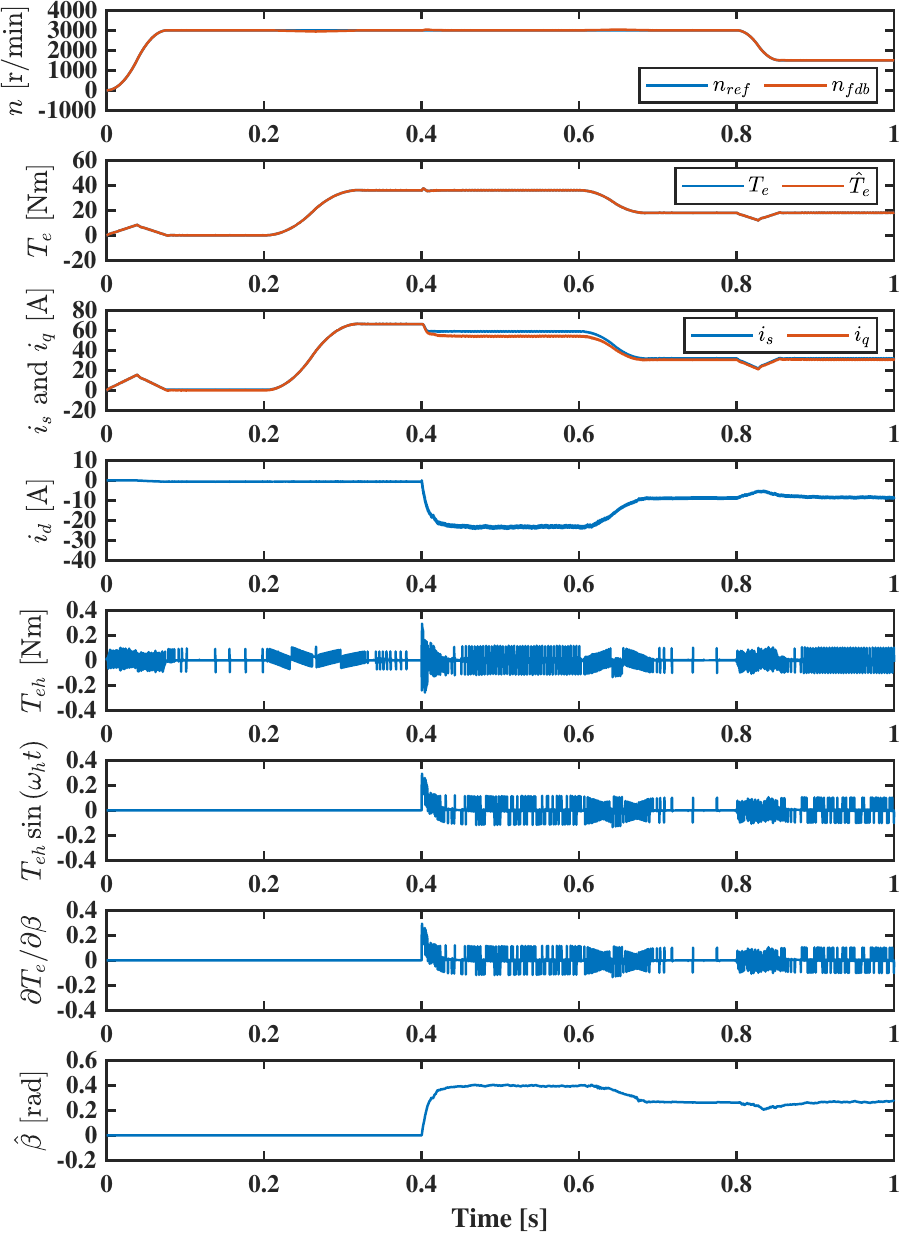}}
                \subfigure[]{\includegraphics[width=0.86\columnwidth]{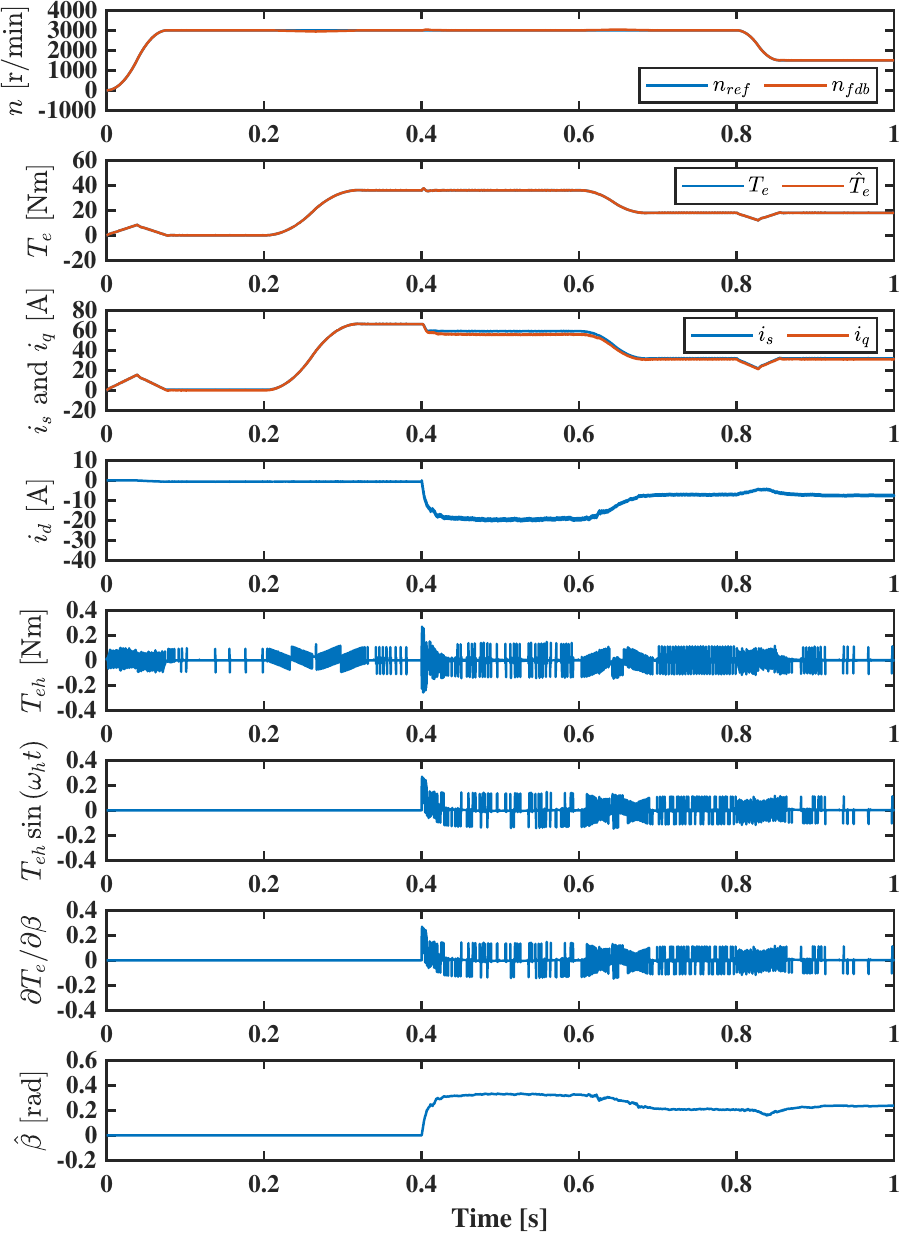}}
			\caption{Simulation results when using extremum seeking. (a) Ideal torque. (b) Observed torque.}
			\label{fig:Waveform of ES}
		\end{figure}

		When using extremum seeking method, injecting a square-wave signal with a frequency of 5 kHz and an amplitude of 0.01 rad into the current vector angle, and an integrator with a gain of 200 is employed. The simulation results when employing ideal torque and observed torque are shown in Fig. \ref{fig:Waveform of ES}. It can be seen that the extremum seeking method works well when using the ideal torque. The current $i_s$ under full load conditions is 66.2 A when using the control strategy of $i_d=0$, and decreases to 58.9 A ($i_d=-23.1$ A and $i_q=54.2$ A) when MTPA is used. Under half load conditions, the currents when using MTPA control are $i_s=31.9$ A, $i_d=-9.3$ A, and $i_q=30.5$ A. Though optimal operating point can be found in the steady state, deviation can be found during the dynamic process when load or speed changes. Better dynamics can be obtained by increasing the amplitude of the injected signal, which, however, leads to larger noise. 
			
\begin{figure} [!t]			
			\centering
                \subfigure[]{\includegraphics[width=0.49\columnwidth]{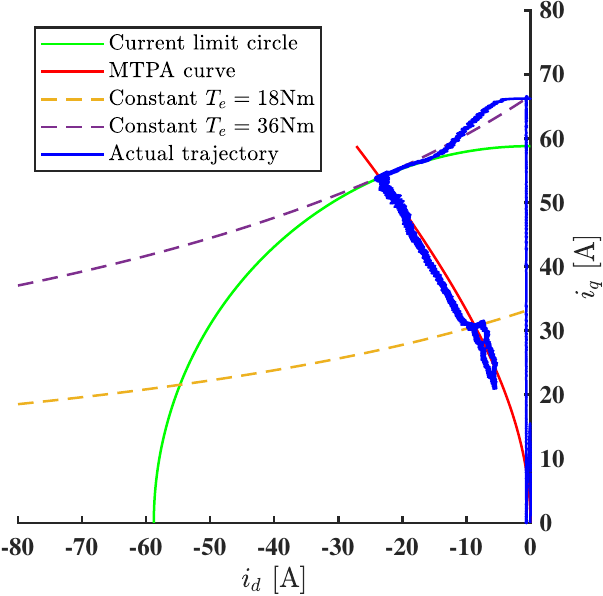}}
			\subfigure[]{\includegraphics[width=0.49\columnwidth]{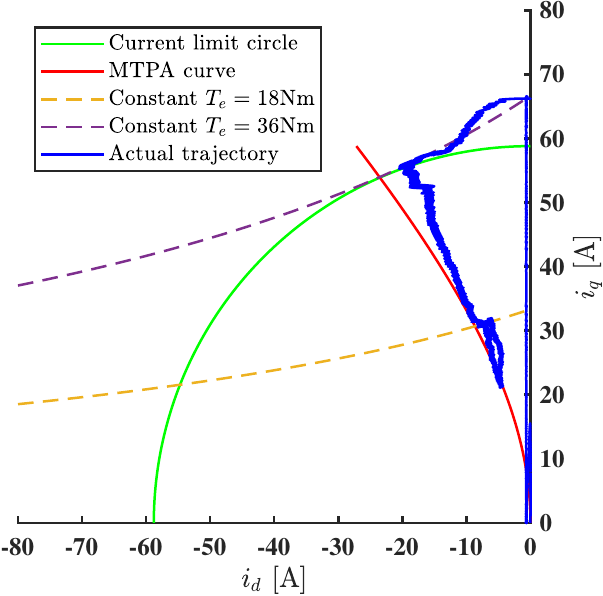}}
			\subfigure[]{\includegraphics[width=0.49\columnwidth]{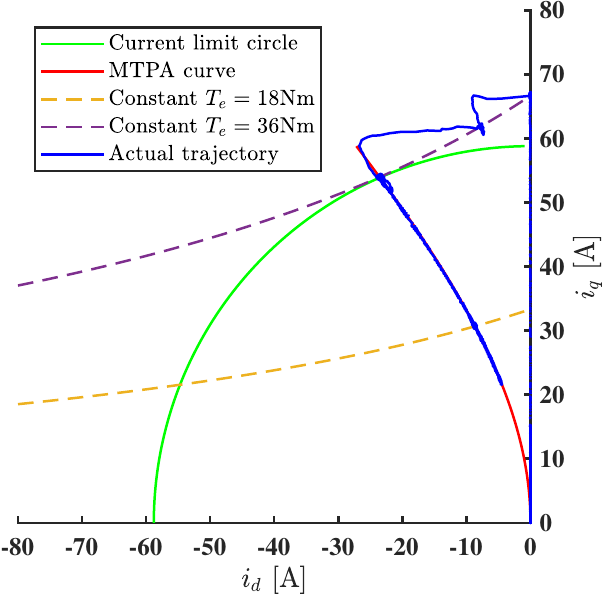}}
			\subfigure[]{\includegraphics[width=0.49\columnwidth]{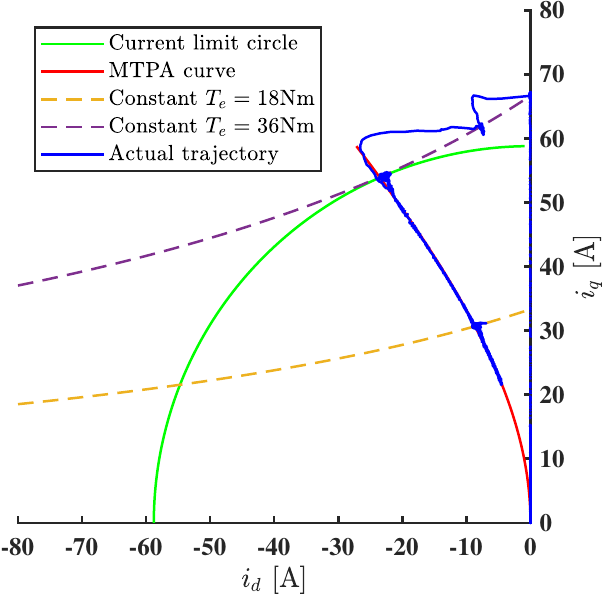}}
            \caption{Current vector trajectory. 
				(a) Extremum seeking with ideal torque. (b) Extremum seeking with observed torque. (c) DCEE with ideal torque. (d) DCEE with observed torque.}
			\label{fig:Trajectory}	
			\vspace{-3ex}		
		\end{figure}

		When employing the observed torque instead of the ideal torque, the dynamic performance becomes poorer, as can see from Fig. \ref{fig:Trajectory}(a) and (b).	
		
		\begin{figure} [htbp]
			\centering
                \vspace{-3ex}
			\setlength{\abovecaptionskip}{-1ex}	
			\subfigure[]{\includegraphics[width=0.86\columnwidth]{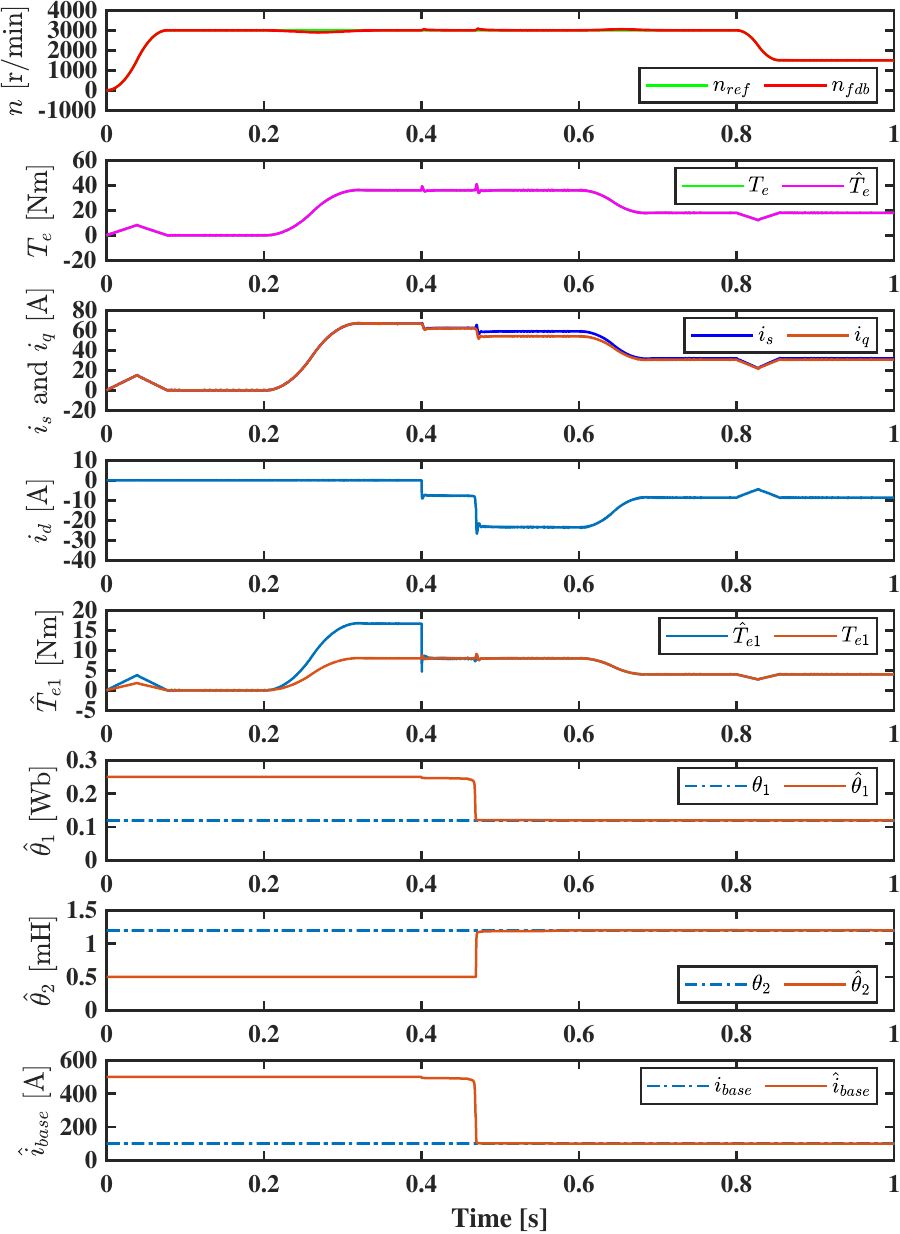}}
                \subfigure[]{\includegraphics[width=0.86\columnwidth]{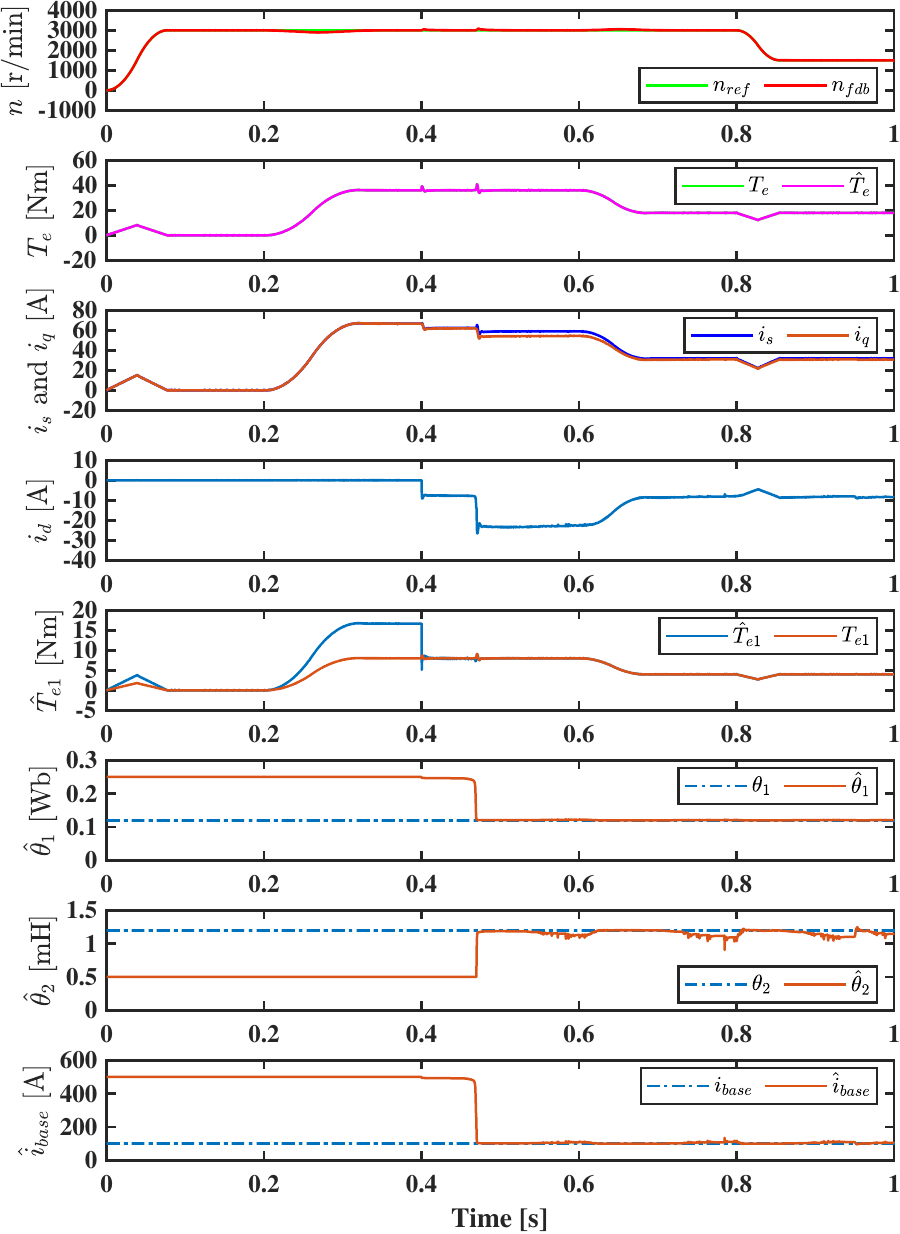}}
			\caption{Simulation results when using DCEE. (a) Ideal torque. (b) Observed torque.}
			\label{fig:Waveform of DCEE}			
		\end{figure}

	\subsection{MTPA Using DCEE}
	
		When using the DCEE method, the forgetting factor in RLS is set as 0.99, simulation results are shown in Fig. \ref{fig:Waveform of DCEE}.
		It can be seen that the proposed DCEE method works well both in the steady state and the transient state. In the time range of 0.2 to 0.4s, the estimated torque $\hat{T}_{e1}$ is calculated using the initial value of the two parameters, i.e., 0.25 Wb and 0.5 mH. At 0.4s, the DCEE algorithm is enabled, the identified permanent magnet flux linkage $\hat{\psi}_f$ converges from its initial value 0.25 Wb to its real value 0.12 Wb, and the identified inductance error $\hat{L}_q-\hat{L}_d$ converges from its initial value 0.5 mH to its real value 1.2 mH. Consequently, the base current $i_{base}$ converges for the initial value changing from 500 to 100. Additionally, it can be seen from Fig. \ref{fig:Trajectory}(c) and (d) that the DCEE method still works well when employing the observed torque instead of the ideal torque.

\section{Conclusion and Future work}

		MTPA control of IPMSM with unknown constant electrical parameters is studied. The conventional extremum seeking method can find the optimal operating point in the steady state, but it deviates from the operating point during the transient process and the situation becomes worse when employing the observed torque instead of the ideal torque. By using the proposed DCEE method, good dynamic performance can be achieved in both steady state and transient state, and it still works well when employing the observed torque instead of the ideal torque. In the future, the proposed DCEE method will be tested in experiments, and IPMSM has variable parameters will be considered.

\bibliographystyle{IEEEtran}	
\bibliography{Bibliography/CDC2024}

\begin{thebibliography}{10}
\providecommand{\url}[1]{#1}
\csname url@samestyle\endcsname
\providecommand{\newblock}{\relax}
\providecommand{\bibinfo}[2]{#2}
\providecommand{\BIBentrySTDinterwordspacing}{\spaceskip=0pt\relax}
\providecommand{\BIBentryALTinterwordstretchfactor}{4}
\providecommand{\BIBentryALTinterwordspacing}{\spaceskip=\fontdimen2\font plus
\BIBentryALTinterwordstretchfactor\fontdimen3\font minus
  \fontdimen4\font\relax}
\providecommand{\BIBforeignlanguage}[2]{{%
\expandafter\ifx\csname l@#1\endcsname\relax
\typeout{** WARNING: IEEEtran.bst: No hyphenation pattern has been}%
\typeout{** loaded for the language `#1'. Using the pattern for}%
\typeout{** the default language instead.}%
\else
\language=\csname l@#1\endcsname
\fi
#2}}
\providecommand{\BIBdecl}{\relax}
\BIBdecl

\bibitem{cai2022critical}
S.~Cai, J.~L. Kirtley, and C.~H.~T. Lee, ``Critical review of direct-drive
  electrical machine systems for electric and hybrid electric vehicles,''
  \emph{IEEE Transactions on Energy Conversion}, vol.~37, no.~4, pp.
  2657--2668, 2022.

\bibitem{rang2004mtpa}
G.~Rang, J.~Lim, K.~Nam, H.-B. Ihm, and H.-G. Kim, ``A mtpa control scheme for
  an ipm synchronous motor considering magnet flux variation caused by
  temperature,'' in \emph{Nineteenth Annual IEEE Applied Power Electronics
  Conference and Exposition, 2004. APEC'04.}, vol.~3.\hskip 1em plus 0.5em
  minus 0.4em\relax IEEE, 2004, pp. 1617--1621.

\bibitem{jung2013current}
S.-Y. Jung, J.~Hong, and K.~Nam, ``Current minimizing torque control of the
  ipmsm using ferrari’s method,'' \emph{IEEE Transactions on Power
  Electronics}, vol.~28, no.~12, pp. 5603--5617, 2013.

\bibitem{morimoto1994effects}
S.~Morimoto, M.~Sanada, and Y.~Takeda, ``Effects and compensation of magnetic
  saturation in flux-weakening controlled permanent magnet synchronous motor
  drives,'' \emph{IEEE Transactions on Industry Applications}, vol.~30, no.~6,
  p. 1632, 1994.

\bibitem{bolognani2010online}
S.~Bolognani, L.~Peretti, and M.~Zigliotto, ``Online mtpa control strategy for
  dtc synchronous-reluctance-motor drives,'' \emph{IEEE Transactions on Power
  Electronics}, vol.~26, no.~1, pp. 20--28, 2010.

\bibitem{sun2015maximum}
T.~Sun, J.~Wang, and X.~Chen, ``Maximum torque per ampere (mtpa) control for
  interior permanent magnet synchronous machine drives based on virtual signal
  injection,'' \emph{IEEE Transactions on Power Electronics}, vol.~30, no.~9,
  pp. 5036--5045, 2015.

\bibitem{sun2016accuracy}
T.~Sun, J.~Wang, and M.~Koc, ``On accuracy of virtual signal injection based
  mtpa operation of interior permanent magnet synchronous machine drives,''
  \emph{IEEE Transactions on Power Electronics}, vol.~32, no.~9, pp.
  7405--7408, 2016.

\bibitem{sun2021extended}
T.~Sun, L.~Long, R.~Yang, K.~Li, and J.~Liang, ``Extended virtual signal
  injection control for mtpa operation of ipmsm drives with online derivative
  term estimation,'' \emph{IEEE Transactions on Power Electronics}, vol.~36,
  no.~9, pp. 10\,602--10\,611, 2021.

\bibitem{zhao2016virtual}
Y.~Zhao, ``Virtual square-wave current injection based maximum torque per
  ampere control for interior permanent-magnet synchronous machines,'' in
  \emph{2016 IEEE Transportation Electrification Conference and Expo
  (ITEC)}.\hskip 1em plus 0.5em minus 0.4em\relax IEEE, 2016, pp. 1--6.

\bibitem{chen2021dual}
W.-H. Chen, C.~Rhodes, and C.~Liu, ``Dual control for exploitation and
  exploration (dcee) in autonomous search,'' \emph{Automatica}, vol. 133, p.
  109851, 2021.

\bibitem{li2022concurrent}
Z.~Li, W.-H. Chen, and J.~Yang, ``Concurrent active learning in autonomous
  airborne source search: Dual control for exploration and exploitation,''
  \emph{IEEE Transactions on Automatic Control}, 2022.

\bibitem{jahns1986interior}
T.~M. Jahns, G.~B. Kliman, and T.~W. Neumann, ``Interior permanent-magnet
  synchronous motors for adjustable-speed drives,'' \emph{IEEE Transactions on
  Industry Applications}, no.~4, pp. 738--747, 1986.

\bibitem{guo1993performance}
L.~Guo, L.~Ljung, and P.~Priouret, ``Performance analysis of the forgetting
  factor rls algorithm,'' \emph{International journal of adaptive control and
  signal processing}, vol.~7, no.~6, pp. 525--537, 1993.

\bibitem{zuo2021two}
Y.~Zuo, J.~Zhu, X.~Yuan, and C.~H. Lee, ``Two-degree-of-freedom quasi-pir
  controller for smooth speed control of permanent magnet vernier machine,'' in
  \emph{2021 IEEE Energy Conversion Congress and Exposition (ECCE)}.\hskip 1em
  plus 0.5em minus 0.4em\relax IEEE, 2021, pp. 5022--5028.

\bibitem{zuo2021digital}
Y.~Zuo, J.~Mei, C.~Jiang, and C.~H. Lee, ``Digital implementation of
  deadbeat-direct torque and flux control for permanent magnet synchronous
  machines in the m--t reference frame,'' \emph{IEEE Transactions on Power
  Electronics}, vol.~36, no.~4, pp. 4610--4621, 2021.

\end{thebibliography}


\end{document}